\documentclass[twocolumn,aps,prb,showpacs,amssymb,amsfonts,superscriptaddress]{revtex4-1}
\usepackage{amsmath}
\usepackage{graphicx}
\usepackage{amssymb}
\usepackage{color}
\usepackage{enumerate}
\usepackage{lipsum}
\usepackage[colorlinks=true,citecolor=blue,linkcolor=blue]{hyperref}

\begin{document}
\title{Anomalous Fermi liquid phase in metallic Skyrmion crystals}
\author{Haruki Watanabe}
\affiliation{Department of Physics, University of California, Berkeley, CA 94720, USA}
\author{S. A. Parameswaran}
\affiliation{Department of Physics, University of California, Berkeley, CA 94720, USA}
\affiliation{Department of Physics and Astronomy, University of California, Irvine, CA 92697, USA}
\author{S. Raghu}
\affiliation{Stanford Institute for Theoretical Physics, Stanford University, Stanford, California 94305, USA}
\affiliation{SLAC National Accelerator Laboratory, 2575 Sand Hill Road, Menlo Park, CA 94025}
\author{Ashvin Vishwanath}
\affiliation{Department of Physics, University of California, Berkeley, CA 94720, USA}
\affiliation{Materials Science Division, Lawrence Berkeley National Laboratories, Berkeley, CA 94720}

\begin{abstract}
In noncentrosymmetric crystals such as MnSi, magnetic order can take the form of a skyrmion crystal (SkX) . In this phase, conduction electrons coupled to the local magnetic moments acquire a Berry phase, leading to an emergent electromagnetism.   Motivated by experimental reports of a non-Fermi-liquid phase in MnSi,  in which resistivity is observed to scale as $\Delta \rho  \sim T^{3/2}$, here we examine the effect of coupling phonons of an incommensurate SkX to electrons. 
Despite the formal similarity to a system consisting of a Fermi surface coupled to an electromagnetic field, the Berry phase fluctuations do not lead to non-Fermi-liquid behavior.   Instead, we propose a different mechanism in which electrons scatter off columnar fluctuation in a three-dimensional SkX. When the effects of lattice-induced anisotropy are neglected, these fluctuations are ultrasoft and  induce an ``anomalous Fermi liquid" in which Landau quasiparticles survive but with an anomalous $\Delta \rho(T)\sim T^{7/4}$ resistivity perpendicular to the columns, and a Fermi-liquid resistivity along them.  \end{abstract}
\maketitle

\section{Introduction}
Recently, much interest has focused on metallic ferromagnets such as MnSi, whose lack of inversion symmetry leads to a twisting of the ferromagnetic order into long period spirals (or helices, hence the term helimagnets). Experiments on the A phase of MnSi~\cite{Pfleiderer2009c,Tokura2012MnSi} have conclusively revealed a crystalline arrangement of spirals with a topological texture---the skyrmion---in each unit cell of a triangular lattice.  Similar skyrmion crystal (SkX) phases have been seen in a host of related materials including FeCoSi~\cite{pfleiderer2010,Tokura2010FeCoSi}, FeGe~\cite{Tokura2010FeGe}, and Cu$_2$OSeO$_3$~\cite{Seki,Adams}. Electrons see the topological texture as a source of magnetic flux, since the spin of the conduction electrons follows the local magnetic moment. The resulting Berry phase is equivalent to the Aharonov-Bohm phase acquired in a magnetic field, and is expected to lead to an additional topological Hall effect~\cite{YBKim,Tokura2013,Binz2008,Binz}. Indeed, precisely such a signal is observed in the skyrmion lattice phase~\cite{Pfleiderer2009c,Lee2009}. Furthermore, the electrons also act on the skyrmion texture, causing it to move when the electrical current exceeds a threshold value. The relatively low threshold currents involved have generated interest for possible spintronic applications~\cite{Rosch,TokuraSpinTorque}. 

While initial observations of the SkX were restricted to the A phase, a small island in the temperature-magnetic field phase diagram, thin films of the same materials have a much expanded region of this phase, extending, for example in the case of FeGe, from zero temperature up to nearly room temperature~\cite{Tokura2013,huang2012, Nagaosa2010}. SkXs have also been observed in single layer iron films deposited on iridium substrate~\cite{IronSkX}, and predicted to occur in ferromagnetic LAO/STO interfaces~\cite{Banerjee2013, Balents2013}. Tantalizing evidence for a 3D skyrmion lattice in powders of MnGe was reported in~\cite{Kanazawa,Kanazawa2011}.

An earlier remarkable experiment~\cite{pfleiderer2001} studied the high-pressure phase of MnSi where spiral magnetic order is lost, and provides one of the best characterized examples of a non-Fermi liquid (nFL). In this phase, the electrical resistivity displays an anomalous temperature dependence \footnote{Here, $\rho(0)$ is the zero-temperature resistivity, which reflects the presence of quenched disorder.} $\Delta \rho \equiv \rho(T) -\rho(0) \sim T^{3/2}$ over nearly three decades in temperature (6 K to 10 mK) and over a wide pressure range. The latter indicates a phase rather than a critical point. Furthermore, the materials themselves are clean with nearly micron-length mean-free paths, and the Fermi-liquid-type resistivity $\Delta \rho \sim T^{2}$ is recovered on exiting this phase by lowering pressure or applying a polarizing magnetic field.  Related phenomenology is seen in the helimagnet FeGe~\cite{Pedrazzini}.  A connection to magnetism was indicated by the unusual ``partial" magnetic order observed in part of this phase, which was interpreted as a spiral with fluctuating orientation \cite{PfleidererPartialOrder,Uemura2007, TurlakovSchmalian2004, Tewari2006}. This triggered early proposals of a magnetic crystal of topological textures~\cite{BinzPRL,Binz2006, Rossler,FischerRosch}.  Further support for a connection between SkXs and nFL transport was provided in recent work which detected a topological Hall effect in the nFL phase~\cite{pfleiderer2013}. 

This motivated us to investigate the fate of low-energy conduction electrons coupled to a skyrmion lattice. We consider an SkX incommensurate with the underlying lattice, which leads, via the Nambu-Goldstone theorem, to low-energy SkX phonon modes.  We further assume that the formation of the SkX leads to Fermi surface reconstruction, and consider coupling electrons near this new Fermi surface to SkX phonons in both 2D and 3D.  
 
Electrons in moving on the Fermi surface are minimally coupled to the Berry vector potential, and the phonon oscillations of the skyrmion crystal induces fluctuations of this gauge field.  Therefore, our problem is very analogous to that of a Fermi sea coupled to a dynamical gauge field.  In the latter case, it is known that gauge bosons are overdamped as a result of interactions with the Fermi surface, and in turn 
electrons show non-Fermi-liquid behavior~\cite{HLR,Stern:1995p1,AltshulerIoffeMillis,NayakWilczekNFL,NayakWilczekNFL2, HolsteinNortonPincus, Nagaosa1992}.  The first half of this paper addresses this issue and proves a negative result: namely, that the unusual electron-phonon coupling does {\it not} lead to non-Fermi-liquid behavior in a SkX, despite the formal similarity with the familiar gauge-field problem. This ultimately stems from the fact that here, the gapless bosons are Goldstone bosons of spontaneously broken translational symmetry.  However a formally similar but distinct situation, where magnetic translation symmetry is spontaneously broken, can lead to the anomalous couplings discussed above~\cite{HarukiAshvin}.
 
The second half of this paper proposes a new mechanism in which anomalous Fermi-liquid behavior arises as a consequence of interactions between electrons and Goldstone bosons.  This requires an ``ultrasoft" Goldstone fluctuation whose dispersion $\epsilon_p\propto p^4$ (depending on the direction of $\vec{p}$), and which may be realized in a columnar skyrmion crystal phase in $3+1$ dimensions.  In this phase, the non-zero winding number of skyrmions gives rise to a term $C(u^x\dot{u}^y-u^y\dot{u}^x)$ in the Goldstone effective Lagrangian [here and below, we assume that the columns are oriented parallel to the $z$ axis, and denote by $\vec{u} = (u_x, u_y)$ the two-dimensional displacement field describing the fluctuations of the columns about equilibrium]. In addition, the elastic energy of the columnar order parameter is known to be anisotropic and lacks the term $(\partial_z\vec{u})^2$~\cite{DeGennes} ---at least when we neglect both the pinning of the columnar axis to the underlying crystal symmetry axes and also possible external magnetic fields.  This form of elastic energy is used in the context of columnar skyrmion crystals in Ref.~\cite{Kirkpatrick2010}, but that work did not take into account the single time derivative term, so that the Goldstone mode had a quadratic, rather than quartic, dispersion in their analysis---and is therefore insufficiently soft to seed non-Fermi-liquid corrections to the quasiparticle relaxation or the transport.  Combining these two properties of a columnar crystal of skyrmions, we demonstrate that the electronic resistivity is proportional to $T^{7/4}$, whereas Goldstone bosons are not overdamped and the electronic quasiparticle lifetime satisfies the Landau criterion that $\omega\tau\rightarrow\infty$ in the limit $\omega\rightarrow 0$. Here, we refer to such examples, in which well-defined quasiparticles  exhibit a resistivity $\rho(T) \propto T^\gamma$ with $\gamma<2$, as {\it anomalous Fermi liquids}. 

This is perhaps a good place to note that although for simplicity we compute a scattering rate for electrons assuming the absence of disorder and/or ``umklapp'' scattering involving large momentum transfers, it is well known that at least one of these must be present in order to permit current relaxation and consequently for the resistivity to have a finite dc ($\omega\rightarrow 0$) limit \footnote{For instance, Ref.~\cite{Kirkpatrick2010} invokes quenched disorder, while a more general discussion on the role of Fermi surface geometry, etc. is provided in ~\cite{MaslovChubukovNFLtransport}.}. Therefore our predictions for non-Fermi-liquid scaling should be viewed more properly as corrections to the finite (disorder-dominated) $T=0$ resistivity $\rho(0)$. Below, we adopt the standard treatment and simply calculate transport relaxation times diagrammatically; the preceding discussion is simply to alert the reader to possible subtleties~\cite{MaslovChubukovNFLtransport} in connecting these results to a measured resistivity.

Potential connections between this mechanism and the observed nFL resistivity is discussed at the end of this paper. An important caveat though is that pinning to crystal axes at low energies will harden the modes responsible for the anomalous Fermi-liquid  behavior. Nevertheless it is interesting to note that a magnetic order with skyrmions naturally leads to a phase with anomalous transport properties but well-defined Landau quasiparticles, thus  adding to the catalog of ``non-Fermi-liquid" phases in three dimensions. 

This paper is organized as follows. In Sec.~\ref{Sec:2} we discuss coupling electrons to the phonons of a SkX and show that, despite the formal similarity to coupling electrons to a gauge field, a careful analysis recovers derivative couplings between electrons and Goldstone modes and Fermi-liquid behavior continues to hold. In Sec.~\ref{Sec:3}, we show that despite derivative couplings, soft columnar fluctuations arising from the breaking of rotation symmetry and the ``Magnus" dynamics of SkX phonons can lead to anomalous transport properties. Finally, in Sec.~\ref{Sec:4}, we discuss  how this mechanism compares to the observations in MnSi, and close with concluding remarks. 

\section{Electrons Coupled to SkX Phonons}
\label{Sec:2}
In this section, we examine the electron-phonon interaction in skyrmion crystals.  As is well known \cite{YBKim,Tokura2013,Binz2008,Binz}, electrons experience a large Berry phase that originates in the non-trivial winding of the background spin texture.  The phonon oscillations of the skyrmion lattice induce fluctuations of the Berry phase, which the electrons experience as a fluctuating gauge field.  Drawing an analogy to the problem of a dynamical gauge field coupled to a Fermi surface, it is natural to expect Landau damping of the Goldstone bosons and non-Fermi-liquid behavior of the electrons.  However, as we now demonstrate, this is {\it not} the case:  Landau damping does not occur, and the Fermi-liquid description remains perfectly adequate.

We begin by considering an electron field $\Psi(\vec{x},t)$, with a spin degree of freedom that interacts with local moments $\vec{n}(\vec{x},t)$ via a strong Hund's rule coupling $J_H$:
\begin{eqnarray}
\Psi_a^\dagger(i\partial_t+\epsilon_{F})\Psi_a-\frac{|\vec{\nabla}\Psi_a|^2}{2m}+\frac{J_H}{2}\vec{n}\cdot\Psi^\dagger\vec{\sigma}\Psi,\label{eq:el0}
\end{eqnarray}
where $\vec{\sigma}$ denotes the vector of Pauli matrices. Here and throughout we set $\hbar=1$. 

\begin{figure}
\begin{center}
\includegraphics[width=\columnwidth]{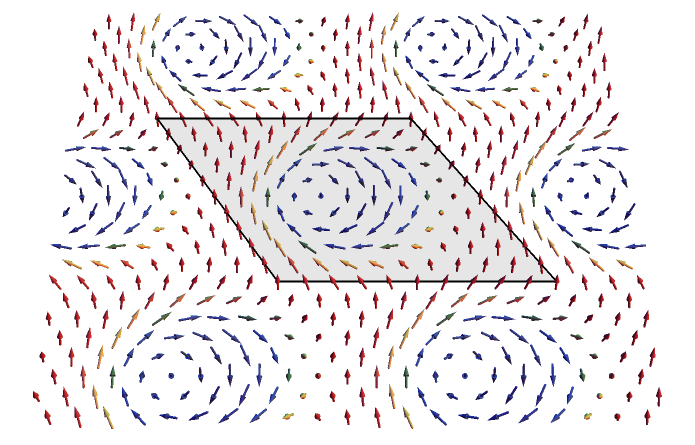}
\end{center}
\caption{(Color online) A skyrmion lattice order of local moments}
\label{fig:skx}
\end{figure}

Suppose the spin texture in the ground state $\vec{n}_0(\vec{x})$ forms a lattice of skyrmions, as illustrated in Fig.~\ref{fig:skx}.  The pitch of the constituent spirals~\cite{BinzPRL, Binz2006} $a_{\text{sk}}$ is set by the Dzyaloshinskii-Moriya coupling $D\vec{n}\cdot\vec{\nabla}\times\vec{n}$ and the SkX is typically incommensurate with the underlying {\it microscopic}\/ lattice, whose  spacing we denote $a_0$.  As a result, Goldstone modes (phonons) that arise from the spontaneously broken translational symmetry are not pinned to the lattice, and thus remain gapless. The low-energy fluctuation of the skyrmion lattice order is dominated by these phonons and we may describe them by introducing a displacement vector field $\vec{u}(\vec{x},t)=(u^x,u^y)$ via
\begin{equation}
\vec{n}(\vec{x},t)=\vec{n}_0(\vec{X}),\quad\vec{X}\equiv\vec{x}-\vec{u}(\vec{x},t).
\end{equation}
Note that all time dependence of $\vec{n}$ is accounted by the displacement field.

\subsection{Local SU(2) rotation}
Since the Hund's rule coupling is strong, it is reasonable to assume that the electron spin is ``slaved" to the local magnetic texture $\vec{n}(\vec{x},t)=\vec{n}_0(\vec{X})$, and to project the electron spin degree of freedom onto this local basis; this projection gives rise to an effective gauge potential acting on the (now ``spinless") electron \cite{Nagaosa2010, Schulz2012}.  We define locally a unitary matrix $U(\vec{x},t)$ by $U^\dagger\,\vec{n}(\vec{x},t)\cdot\vec{\sigma}\,U=\sigma_3$ or
\begin{eqnarray}
U(\vec{x},t)=\label{eq:Undef}
\begin{pmatrix}
\cos\frac{\theta}{2}&-e^{-i\phi}\sin\frac{\theta}{2}\\
e^{i\phi}\sin\frac{\theta}{2}&\cos\frac{\theta}{2}\\
\end{pmatrix}.
\end{eqnarray}
where $(\theta(\vec{x},t),\phi(\vec{x},t))$ is the spherical coordinate of $\vec{n}(\vec{x},t)$, and rewrite the Lagrangian in terms of $\tilde{\Psi}=(\tilde{\Psi}_1,\tilde{\Psi}_2)\equiv U^{-1}\Psi$.  We note that $(U,\tilde{\Psi})$ and $(Ue^{i\Lambda(\vec{x},t)\sigma_3},e^{-i\Lambda(\vec{x},t)\sigma_3}\tilde{\Psi})$ should be equivalent since they describe the same $\vec{n}(\vec{x},t)$ and $\Psi(\vec{x},t)$.  There is thus a gauge redundancy and we will fix the gauge shortly.

As a consequence of the local rotation, the kinetic term of the electron acquires an SU(2) Berry vector potential $-iU^\dagger\partial_\mu U$. The original Hund's rule coupling becomes a constant chemical potential\footnote{Observe that we absorb the strength of the local moment into the definition of $J_H$ so that $\hat{n}$ is a unit vector (as should be evident from Eq.(\ref{eq:Undef}).)} $J_H\tilde{\Psi}^\dagger\sigma_3\tilde{\Psi}/2$.  The spin parallel component $\psi\equiv\tilde{\Psi}_1$ has a lower energy than that of the antiparallel component $\tilde{\Psi}_2$ by $J_H$.  Assuming $J_H$ is large enough, we may integrate out $\tilde{\Psi}_2$ to get a low-energy theory in terms of $\psi$.  

After these manipulations, the electron Lagrangian~\eqref{eq:el0} now reads
\begin{eqnarray}
\psi^\dagger(i\partial_t+\epsilon_{F}-\mathcal{A}_t-V)\psi-\frac{|(\vec{\nabla}-i\vec{\mathcal{A}})\psi|^2}{2m}.\label{eq:el1}
\end{eqnarray}
Here, \begin{equation}
V(\vec{x},t)\equiv\frac{[\partial_i\vec{n}_0(\vec{X})]^2}{8m}
\end{equation}
is a periodic lattice potential and $\mathcal{A}_\mu$ is the U(1) Berry vector potential including phonon fluctuations. In the ground state $\vec{u}=0$, $\vec{\mathcal{A}}(\vec{x},t)$ reduces to a static vector potential,
\begin{equation}
\vec{A}(\vec{x})\equiv\frac{\cos\theta_0(\vec{x})-1}{2}\vec{\nabla}\phi_0(\vec{x})
\end{equation}
and the associated magnetic field coincides with the number density of skyrmions (upto a factor of $-2\pi$):
\begin{equation}
B_0(\vec{x})=[\vec{\nabla}\times\vec{A}(\vec{x})]_z=-\frac{1}{2}\vec{n}_0(\vec{x})\cdot\partial_x\vec{n}_0(\vec{x})\times\partial_y\vec{n}_0(\vec{x}).
\end{equation}
The full time-dependent $\mathcal{A}$ can be expressed in terms of its ground-state value $\vec{A}$ and the field $\vec{X}$ as
\begin{eqnarray}
\vec{\mathcal{A}}(\vec{x},t)=A^i(\vec{X})\vec{\nabla}X^i,\,\,\mathcal{A}_t(\vec{x},t)=-A^i(\vec{X})\partial_tX^i\label{mA}\label{berry}.
\end{eqnarray}

Note that the effective Lagrangian Eq.~\eqref{eq:el1} has exact translational symmetry,
\begin{eqnarray}
\vec{u}'(\vec{x}+\vec{\epsilon},t)=\vec{u}(\vec{x},t)+\vec{\epsilon},\quad\psi'(\vec{x}+\vec{\epsilon},t)=\psi(\vec{x},t).\label{translation}
\end{eqnarray}

While $B_0(\vec{x})$ always has a spatially oscillating component for any noncollinear spin texture, a spatially uniform component $\bar{B}_0$ exists only when the total number of skyrmions is nonzero---as in a SkX.  We assume $\bar{B}_0>0$ and use the notation $\vec{B}_0=B_0\hat{z}$ in the following.

As the phonon oscillates, so does the effective gauge field $\mathcal{A}_{\mu}$ and as a consequence the electrons experience a fluctuating electromagnetic field,
\begin{eqnarray}
\vec{E}(\vec{x},t)=-\vec{\nabla}\mathcal{A}_t-\partial_t\vec{\mathcal{A}}=\vec{B}_0(\vec{x})\times\vec{v}(\vec{x},t),\\
\vec{B}(\vec{x},t)=\vec{\nabla}\times\vec{\mathcal{A}}=\vec{B}_0(\vec{x})\mathrm{det}(\partial X^j/\partial x^i),\label{magnetic}
\end{eqnarray}
where $\vec{v}\equiv \vec{J}/J^t=\partial_t\vec{u}+O(u^2)$ and $J^\mu=(1/8\pi)\epsilon^{\mu\nu\lambda}\vec{n}\cdot\partial_\nu\vec{n}\times\partial_\lambda\vec{n}=-\epsilon^{\mu\nu\lambda}\partial_\nu \mathcal{A}_\lambda$ is the conserved current of skyrmions.  In this fashion, the electron-phonon interaction is built into the Lagrangian Eq.~\eqref{eq:el1}. Before discussing its effect, let us first summarize the free electron wave functions.

\subsection{Electron Dispersion in SkX}
By setting $\vec{u}=0$ (or $\vec{X}=\vec{x}$), we find the free Lagrangian of electrons,
\begin{eqnarray}
\psi^\dagger(i\partial_t+\epsilon_{F}-V)\psi-\frac{|(\vec{\nabla}-i\vec{A})\psi|^2}{2m}.\label{eq:elfree}
\end{eqnarray}
The periodic potential $V(\vec{x})=[\partial_i\vec{n}_0(\vec{x})]^2/(8m)$ satisfies $V(\vec{x}+\vec{a}_1)=V(\vec{x}+\vec{a}_2)=V(\vec{x})$ where $\vec{a}_{1,2}$ are primitive lattice vectors of the SkX.

The aforementioned gauge redundancy (parametrized by $\Lambda$ above) allows us to choose an arbitrary gauge potential as long as it describes the same effective magnetic field.  We abandon the periodicity of $\vec{A}$ and choose the Landau gauge over the entire system,
\begin{equation}
\vec{A}(\vec{x})=\bar{B}_0\begin{pmatrix}-y\\0\end{pmatrix}+\delta\vec{A}(\vec{x}),
\end{equation}
where the first term describes the uniform component of the magnetic field and the periodic vector potential $\delta\vec{A}(\vec{x}+\vec{a}_1)=\delta\vec{A}(\vec{x}+\vec{a}_2)=\delta\vec{A}(\vec{x})$ is for the periodically oscillating component.  (Alternatively, we could retain a periodic $\vec{A}$ by introducing at least two patches per unit cell.)

The detailed form of the periodic potential $V(\vec{x})$ and the periodic magnetic field $\vec{\nabla}\times\delta\vec{A}(\vec{x})$ is not important, but in their absence, the Lagrangian~\eqref{eq:elfree} can describe only {\it flat}\/ Landau levels.  In order to get a Fermi surface, therefore, we have to take into account at least one of these periodic effects to give rise to a nonzero dispersion to the band structure.  

In the single-particle picture, the Hamiltonian corresponding to the Lagrangian~\eqref{eq:elfree}) can be expressed as
\begin{eqnarray}
H_0=\frac{\vec{\pi}^2}{2m}+V(\vec{x})-\epsilon_F,\quad \vec{\pi}=\vec{p}-q\vec{A}(\vec{x}).\label{eq:Helfree}
\end{eqnarray}
This Hamiltonian commutes with the lattice translation $T_1=e^{i\vec{P}\cdot\vec{a}_1}$ and $T_2=e^{i\vec{P}\cdot\vec{a}_2}$ where $\vec{P}=(p^x,p^y+q\bar{B}_0x)$ is the magnetic translation. $T_1$ and $T_2$ also commute since the flux per a unit cell is precisely $q\bar{B}\vec{z}\cdot\vec{a}_1\times\vec{a}_2=2\pi$.  We can thus simultaneously diagonalize $H_0$ and $T_{1,2}$; the corresponding eigenstates $|n\vec{k}\rangle$ have eigenvalues $\epsilon_{n\vec{k}}$ and $e^{i\vec{k}\cdot\vec{a}_i}$ for $H_0$ and $T_i$, respectively.  We expand the electron field $\psi(\vec{x},t)$ as
\begin{eqnarray}
&\psi(\vec{x},t)=\sum_{n\vec{k}} \psi_{n\vec{k}}(\vec{x})c_{n\vec{k}}(t),&\\
&\psi_{n\vec{k}}(\vec{x})=\langle\vec{x}|n\vec{k}\rangle,\quad\{c_{n\vec{k}},c_{n'\vec{k}'}^\dagger\}=\delta_{\vec{k},\vec{k}'}\delta_{n,n'}.&
\end{eqnarray}

Although $\psi_{n\vec{k}}(\vec{x})$ is not  a plane wave, we may still use some useful properties of Bloch states. (See Ref.~\cite{haldane} for the explicit formula for $\psi_{n\vec{k}}(\vec{x})$.) For example, we have
\begin{eqnarray}
&\langle n'\vec{k}'|e^{i\vec{q}\cdot\vec{x}}|n\vec{k}\rangle\propto \delta_{\vec{q},\vec{k}'-\vec{k}+\vec{G}},&\\
&\langle n\vec{k}|\vec{j}_{\text{el}}|n\vec{k}\rangle=\vec{\nabla}_{\vec{k}}\epsilon_{n\vec{k}},&
\end{eqnarray}
since these relations are solely based on the definition $H_0|n\vec{k}\rangle=\epsilon_{n\vec{k}}|n\vec{k}\rangle$ and $T_{i}|n\vec{k}\rangle=e^{i\vec{k}\cdot\vec{a}_{i}}|n\vec{k}\rangle$.  We can write the Lagrangian $L_0=\int\mathrm{d}^dx\mathcal{L}_0$ as
\begin{equation}
L_0=\sum_{n\vec{k}}c_{n\vec{k}}^\dagger(i\partial_t-\epsilon_{n\vec{k}})c_{n\vec{k}}.
\end{equation}
This clarifies that the electron Green's function is still $G_0^{-1}=\omega-\epsilon_{n\vec{k}}$.

\begin{figure}
\begin{minipage}[t]{0.39\columnwidth}
\begin{center}
\includegraphics[clip, width=0.99\columnwidth]{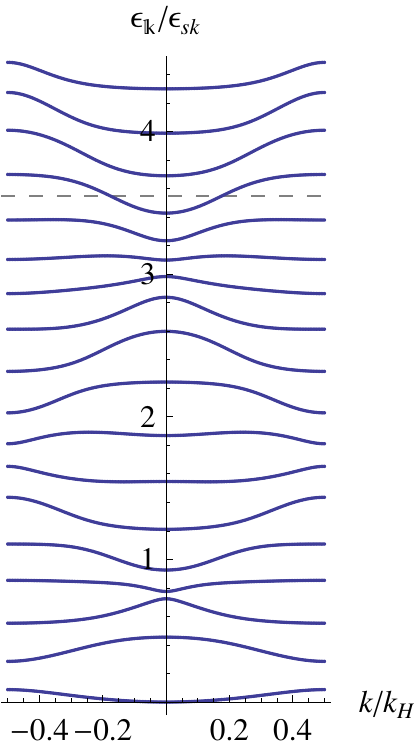}
\end{center}
\end{minipage}
\begin{minipage}[t]{0.59\columnwidth}
\begin{center}
\includegraphics[clip, width=0.99\columnwidth]{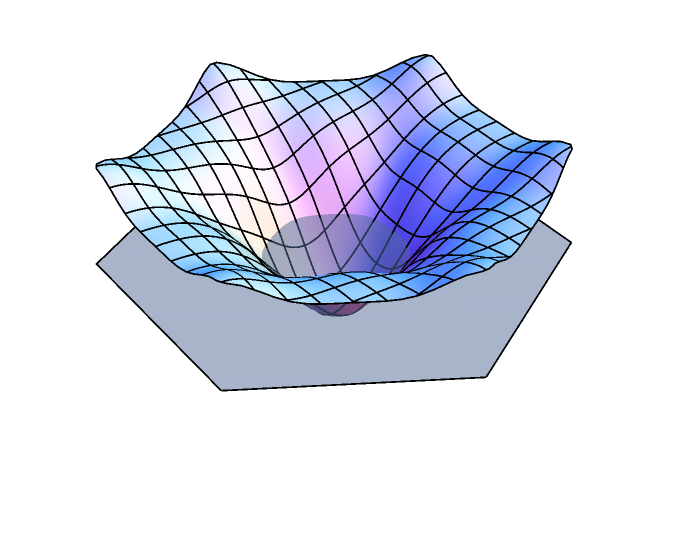}
\end{center}
\end{minipage}
\caption{(Color online) Left: The low-energy electronic band structure of the model (\ref{eq:el0}) for $J_H=10(k_H^2/2m)$.   Right:  Dispersion along $\vec{k}=\vec{G}_1$, with the dashed line indicating the Fermi energy. }
\label{fig:electronb}
\end{figure}

Figure~\ref{fig:electronb} shows the  band structures obtained by numerically solving the Bloch problem of the Lagrangian (\ref{eq:el0}) for a given skyrmion lattice configuration $\vec{n}_0(\vec{x})$.  
  The original electron bands with  typical wave number $\pi/a_0$ are folded many times and reconstruct a smaller Fermi surface within the reduced Brillouin zone of the SkX $\sim \pi/a_{\text{sk}}$.   Interband separations are the order of $\omega_c\equiv q\bar{B}_0/m$ and each band has the Chern number $+1$.

We assume that the conduction electrons fill a number of low-lying bands and leave a partially filled band with electronic density $n_e \ll n^{(0)}_e$
~\footnote{
We neglect the filled bands as they are gapped and thus irrelevant to the low-energy dynamics.
More precisely, integrating out filled bands generates a Chern-Simons term $\frac{\sigma_{xy}}{2}\epsilon^{\mu\nu\lambda}(a_\mu+\mathcal{A}_\mu)\partial_\nu(a_\lambda+\mathcal{A}_\lambda)$, where $a_\mu$ is the real U(1) gauge field.  This term encodes the fact that Skyrmions have electric charge $\frac{\sigma_{xy}}{e}$ and interact with each other through the long-range Coulomb potential.  However, conduction electrons in the partially filled band screen the interaction, making it short-ranged. 

There is yet another subtlety.  To be consistent with our neglect of the spin-antiparallel component $\tilde{\Psi}_2$, the new Fermi level must be below $J_H$ as measured from the bottom of the lowest electron band. Thus, the number of filled bands, which is typically the order of $(a_{\text{sk}}/a_0)^2$, should be less than $J_H/\omega_c$. This requires $J_H>(4\pi/\sqrt{3})(ma_0^2)^{-1}$.  We assume this condition for simplicity, but we caution that this is a much more severe condition than the naive one, $J_H\gg \frac{v_F^{(0)}}{a_{\text{sk}}}\approx\frac{1}{ma_0a_{\text{sk}}}$.
}.  We focus on the partially filled band and neglect all others.  Our interest is in the fate of this Fermi surface --- whether the electron-phonon interaction leads to non-Fermi-liquid behavior or not.

\subsection{Electron-phonon interaction}
We expand the Berry gauge field $\mathcal{A}_\mu(\vec{x},t)$ corresponding to a phonon fluctuation in a power series in $\vec{u}$:
\begin{eqnarray}
\mathcal{A}_t&=&-\frac{1}{2}\vec{B}_0(\vec{x})\cdot\vec{u}\times\dot{\vec{u}}-\partial_t\chi+O(u^3),\\
\vec{\mathcal{A}}&=&\vec{A}(\vec{x})-\vec{B}_0(\vec{x})\times\vec{u}+\frac{1}{2}(\vec{u}\cdot\vec{\nabla})\vec{B}_0(\vec{x})\times\vec{u}\notag\\
&&+\frac{1}{2}B_0(\vec{x})\epsilon_{ij}u^i\vec{\nabla} u^j+\vec{\nabla}\chi+O(u^3),
\end{eqnarray}
where $\chi(\vec{x},t)=-\vec{u}\cdot\vec{A}(\vec{x})-(1/4)u^iu^j(\partial_iA_j+\partial_jA_i)(\vec{x})+O(u^3)$. We remove $\partial_\mu\chi$ by a gauge transformation. This may change the transformation rule~\eqref{translation} but it does not affect the following analysis.  

The linear $\vec{u}$ term in the expansion can be understood intuitively in the limit of a uniform magnetic field $B_0(\vec{x})=B_0$.  When the lattice uniformly expands, the magnetic field should decrease as $B_0\rightarrow B_0(1-\vec{\nabla}\cdot\vec{u})$ as described by Eq.~\eqref{magnetic}.   To reproduce this change, $\vec{\mathcal{A}}$ must contain a term $-\vec{B}_0\times\vec{u}$.  

\begin{figure}
\begin{center}
\includegraphics[width=0.4\columnwidth,clip]{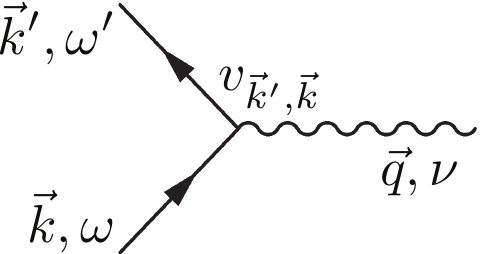}
\end{center}
\caption{The electron-phonon vertex with one external phonon line.}
\label{fig:vertex}
\end{figure}

The periodic potential can also be expanded as
\begin{equation}
V(\vec{X})=V(\vec{x})-\vec{u}\cdot\vec{\nabla} V(\vec{x})-\frac{\partial_i\vec{n}\cdot\partial_j\vec{n}}{4m}\partial_iu^j+O(u^2).
\end{equation}
Therefore, the electron-phonon interaction is given by $\mathcal{L}_{\text{int}}=\mathcal{L}_{\text{int}}^{(1)}+\mathcal{L}_{\text{int}}^{(2)}+O(u^2)$, where
\begin{eqnarray}
\mathcal{L}_{\text{int}}^{(1)}&=&-q\vec{B}_0(\vec{x})\cdot\vec{u}\times\vec{j}_{\text{el}},\label{int1}\\
\mathcal{L}_{\text{int}}^{(2)}&=&\left[\vec{u}\cdot\vec{\nabla} V(\vec{x})+\frac{\partial_i\vec{n}\cdot\partial_j\vec{n}}{4m}\partial_iu^j\right]\psi^\dagger\psi,\label{int2}
\end{eqnarray}
and express these interactions into the form (see Fig.~\ref{fig:vertex}),
\begin{equation}
L_{\text{int}}=\sum_{\vec{k}',\vec{k}}\vec{v}_{\vec{k}',\vec{k}}\cdot\vec{u}_{\vec{q}}\psi^\dagger_{\vec{k}'}\psi_{\vec{k}}\label{vertex}
\end{equation}
where $\vec{q}=\vec{k}'-\vec{k}$ describes the momentum transfer under the scattering and $\vec{v}_{\vec{k}',\vec{k}}=(\vec{v}_{\vec{k},\vec{k}'})^*$ is called a (bare) vertex function. 

The coupling in Eq.~\eqref{int1}, which is first discussed in Ref.~\cite{Nagaosa2DSkX}, is singular in the sense that no derivatives act on the displacement field $\vec{u}$ and is reminiscent of the interaction $e\vec{A}\cdot\vec{j}_{\text{el}}$ in the well-studied problem of the dynamical gauge field interacting with a Fermi surface.  Indeed, the contribution of $\mathcal{L}_{\text{int}}^{(1)}$ to $\vec{v}_{\vec{k},\vec{k}}$ is
\begin{equation}
\vec{v}_{\vec{k},\vec{k}}^{(1)}\approx e\bar{B}_0\hat{z}\times \vec{\nabla}_{\vec{k}}\epsilon_{\vec{k}}.\label{vertexskx}
\end{equation}
Here we neglected the umklapp scattering by the periodic part of the magnetic field.  The vertex (\ref{vertexskx}) looks similar to the corresponding one for the real gauge field problem, $e\vec{k}/m=e\vec{\nabla}_{\vec{k}}\epsilon_{\vec{k}}$.  The significance of these vertices is that they do not vanish even in the limit of small energy-momentum transfer $\vec{k}'\rightarrow \vec{k}$, indicating a strong coupling between electron excitations and low-energy Goldstone bosons.  Therefore, reasoning by analogy we naively expect the standard non-Fermi-liquid behavior to emerge also in this electron-phonon coupled system.  

However, this discussion omitted another term that also contains $\vec{u}$ without derivatives: namely, 
$\vec{u}\cdot\vec{\nabla}V(\vec{x})\psi^\dagger\psi\in\mathcal{L}_{\text{int}}^{(2)}$. 
This term always exists in electron-phonon coupled systems and usually simply produces corrections within the Fermi-liquid theory.
Here, however, its role is crucial in invalidating the naive expectation above.  Note also that in the absence of the periodic potential $V=0$, it is the periodicity in the magnetic field that produces a nonzero electronic dispersion. In this case, neglecting the umklapp process can no longer be justified, as it would be inconsistent with the preceding analysis.

To simply this complicated situation and unambiguously discuss the scattering in the limit of $\vec{k}'\rightarrow\vec{k}$, it is convenient to express the electron-phonon interaction in the singe-particle picture in the form~\footnote{We thank Max Metlitski for useful discussion.}
\begin{eqnarray}
H_{\text{int}}=\vec{u}\cdot\left(e\vec{B}_0(\vec{x})\times\vec{j}_{\text{el}}+\vec{\nabla}V(\vec{x})\right)=\vec{u}\cdot[i\vec{\pi},H_0],
\end{eqnarray}
where $\vec{j}_{\text{el}}=\vec{\pi}/m$ and $H_0$ are defined in Eq.~\eqref{eq:elfree}.  Since $|\vec{k}\rangle$ is an eigenstate of $H_0$, 
\begin{eqnarray}
\langle\vec{k}'|H_{\text{int}}|\vec{k}\rangle&=&\vec{u}\cdot\langle\vec{k}'|[i\vec{\pi},H_0]|\vec{k}\rangle\notag\\
&=&\vec{u}\cdot\langle\vec{k}'|\vec{\pi}|\vec{k}\rangle i(\epsilon_{\vec{k}}-\epsilon_{\vec{k}'})
\end{eqnarray}
for a constant $\vec{u}$. Thus, as a whole, the electron-phonon vertex vanishes in the limit $\vec{k}'\rightarrow\vec{k}$.  This is to some extend expected since a constant displacement should not have a physical effect on electron states.  However, this conclusion is not quite trivial, since there is the well-known example of nematic Fermi liquids, where scattering of electrons by the orientational Goldstone modes does not vanish even in the limit of small energy-momentum transfer~\cite{Oganesyan2001}, due to the singularity in the matrix element.

In summary, electron-phonon interactions in the SkX become weaker and weaker in the long-wavelength limit and there is no qualitative difference from the usual electron-phonon problem, despite the apparent analogy to a Fermi sea minimally coupled  to a fluctuating gauge field.  Therefore, the consequence of the electron-phonon interaction must be the same as the well-studied electron-phonon problem, except that phonons in SkX have a quadratic dispersion due to the non-zero winding number of the spin configuration.  We may understand this as a consequence of a cancellation between contributions from the two linear-$\vec{u}$ terms that couple to the current and density, respectively.

To estimate the lifetime of the quasiparticles, it is useful to use the Fermi golden rule
\begin{equation}
\frac{1}{\tau} = 2\pi\sum_{\text{final}} |\langle {\text{final}} | \,H_{\text{int}} |{\text{initial}}\rangle|^2\delta(E_f-E_i).\label{golden}
\end{equation}
Consider an energy $\omega$ electron above the Fermi surface, which decays to a lower energy state by emitting a SkX phonon. Due to the derivative coupling, the matrix element scales as $q^2$. Since the electrons are much faster than the phonons, their scattering is essentially in the direction orthogonal to their initial momentum, and the maximum momentum deviation that can be supplied by a SkX phonon is $q \sim \sqrt{\omega}$, owing to the quadratic dispersion. This yields a lifetime 
\begin{equation}
1/\tau \propto q^3\propto\omega^{3/2}
\end{equation}
and therefore, the single-particle electron excitation become sharper and sharper as $\omega\rightarrow 0$. The transport relaxation time, evaluated by adding a factor of $(1-\cos\theta_{\vec{k}',\vec{k}})$ in the calculation of the self-energy, behaves as $\tau_{\mathrm{tr}}^{-1}\propto T^{5/2}$. Therefore, electrons can still be well-described by Landau's Fermi-liquid theory.  Phonons are not overdamped either; the singular correction $\propto -i\nu/q$, which usually leads to the Landau damping of bosons as $q\rightarrow 0$, is canceled by a $q^2$ factor coming from the vertex function.  

\section{Anomalous Fermi Liquid from Columnar Fluctuations}
\label{Sec:3}
In this section, we discuss a new type of non-Fermi scaling of the  resistivity that arises through the interaction between Goldstone modes and a Fermi surface.  We clarified in the previous section that electron-phonon interactions vanish in the limit of small energy-momentum transfer. Such weak couplings (or {\it derivative}\/ couplings) are usually irrelevant perturbations.  However, the Goldstone mode we discuss in this section has an extraordinarily soft dispersion, which changes the scaling law of the boson field.  As we shall see below, although electrons receive a large self-energy correction due to this ``ultrasoft" Goldstone mode, the Landau quasiparticle remains well defined; {\it i.e.}, the quasiparticle lifetime $\tau$ obeys $\omega\tau \rightarrow \infty$ as $\omega\rightarrow 0$. However, the scattering of quasiparticles off the Goldstone modes {\it does} lead to anomalous conductivity: the transport relaxation time $\tau_\text{tr}\sim T^{7/4}$ at low temperatures, dominating the quadratic temperature dependence expected for a conventional Fermi liquid. Meanwhile, the Goldstone bosons are not overdamped as the electron-phonon vertex function vanishes in the limit of $\vec{k}'=\vec{k}$. For the sake of clarity, and to emphasize the distinction between this example and a true non-Fermi liquid for which there are no well-defined quasiparticles, we will refer to this as an ``anomalous Fermi liquid."

\begin{figure}
\begin{center}
\includegraphics[width=0.4\columnwidth,clip]{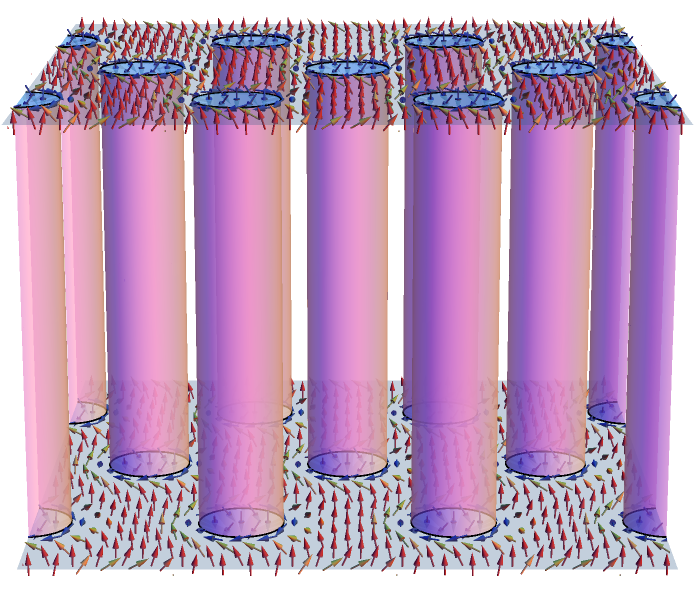}
\end{center}
\caption{(Color online) A 3D columnar SkX. The columnar order does not break the translational symmetry in the $z$ direction, while each $x$-$y$ plane supports, for example, a hexagonal 2D SkX. The finite skyrmion density leads to linear time derivative dynamics for the SkX phonons and in the absence of lattice anisotropy, the spontaneous breaking of 3D rotation symmetry leads to anomalously soft phonons with dispersion $\omega \sim (k_x^2+k_y^2)+k_z^4$.}
\label{fig:3dcolumnar}
\end{figure}

Here we will consider a 3D columnar order of skyrmions, as illustrated in Fig.~\ref{fig:3dcolumnar}, ignoring the effects of the crystal lattice so that we begin with continuous translation and rotation symmetry of isotropic 3D space. Also, we assume there is no applied magnetic field. Suppose now that a skyrmion crystal is formed, such that skyrmions form a hexagonal columnar lattice which breaks rotation symmetry by spontaneously choosing a ``$z$ axis" along which the columns are directed, as well as translation symmetry in the orthogonal ``$x$-$y$" plane. 

The elastic energy associated to the lattice displacement $\vec{u}=(u^x,u^y)^T$ is given by
\begin{equation}
\mathcal{E}_{\text{el}}=\frac{\rho_1}{2}\left(\nabla_{\perp}\cdot\vec{u}\right)^2+\frac{\rho_2}{2}\left(\nabla_{\perp}\times\vec{u}\right)^2+\frac{\rho_z}{2}(\partial_z^2\vec{u})^2
\end{equation}
Crucially, this expression lacks a $(\partial_z \vec{u})^2$ term.  This is only true when we neglect both pinning to crystal axes and possible external fields that may stabilize the columnar order; to avoid lengthy digressions to consider these subtleties,  here we content ourselves with considering the theoretical situation of  columnar order in the {\it free}\/ space without any external fields.

Recalling that the phonon dynamics will be governed by the Berry phase term due to the nonzero skyrmion density, we obtain the effective Lagrangian for phonon fluctuations~\cite{MacDonaldQHSkX,Nagaosa2DSkX},
\begin{equation}
\mathcal{L}_{\text{ph}}=s B_0(u_x\dot{u}_y-u_y\dot{u}_x)-\mathcal{E}_{\text{el}}.
\end{equation}
Here, $s=S/a_0^3$ where $a_0$ is the lattice constant of the microscopic lattice and $S$ is the spin at each site.
The electronic Lagrangian is assumed to be the same as \eqref{eq:elfree} with the the addition of a extra spatial derivative term along $z$.

The phonon Green's function is
\begin{equation}
D_0=\frac{\epsilon_{\vec{q}}\sigma_0-\nu\sigma_2}{2sB_0(\nu+i\delta-\epsilon_{\vec{q}})(\nu-i\delta+\epsilon_{\vec{q}})},
\end{equation}
where we set $\rho_1=\rho_2=\rho_{\parallel}$ for simplicity and $\epsilon_{\vec{q}}=(2sB_0)^{-1}(\rho_{\parallel}q_{\parallel}^2+\rho_z q_z^4)$.

As we discussed in the previous section, electron-phonon interactions must vanish in the limit of small energy-momentum transfer.  We may therefore expand the corresponding vertex as a series in $\vec{q}=\vec{k}'-\vec{k}$:
\begin{equation}
v^i_{\vec{k}',\vec{k}}=iV^i_jq^j+O(q^2).
\end{equation}
Here, $V_j^i$ ($i=x,y$, $j=x,y,z$) is a real constant. Due to this linear $q$ dependence of $v^i_{\vec{k}',\vec{k}}$, the singular correction $-i\nu/q$ is smeared and the boson self-energy correction is subleading compared to the bare propagator.

Now let us evaluate the electron self-energy.  For $\omega>0$, its imaginary part is given by
\begin{eqnarray}
\tau^{-1}&\equiv&-2\text{Im}\Sigma(\vec{k},\omega)\notag\\
&=&\int\frac{\mathrm{d}^3q}{(2\pi)^3}\frac{V^i_jq^jV^i_kq^k}{4sB_0}f(\xi_{\vec{k}'})\delta(\omega-\xi_{\vec{k}'}+\epsilon_{\vec{q}}),
\end{eqnarray}
where $\xi_{\vec{k}}=k_{\parallel}^2/2m_{\parallel}+k_z^2/2m_z-\mu$ is the electron excitation energy and $f(\omega)=\theta(-\omega)$ is the Fermi distribution function.

In particular, for $\vec{k}=k_F\hat{x}$, we first perform the $q_x$ integral (perpendicular to the Fermi surface) and find
\begin{eqnarray}
\tau^{-1}&=&\frac{V^i_zV^i_z}{4sB_0}\int\frac{\mathrm{d}^3q}{(2\pi)^3}q_z^2\theta(\omega-\epsilon_{\vec{q}})\notag\\
&&\times\delta(\omega-(2k_Fq_x+q_y^2)/2m_{\parallel}-q_z^2/2m_z+\epsilon_{\vec{q}})\notag\\
&\simeq&\frac{V^i_zV^i_zm_{\parallel}}{4sB_0k_F}\int\frac{\mathrm{d}q_y\mathrm{d}q_z}{(2\pi)^3}q_z^2\theta\Big(\omega-\frac{\rho_{\parallel}q_y^2+\rho_z q_z^4}{2sB_0}\Big).
\end{eqnarray}
In the above calculation, we neglect terms in the integrand of the form $q_i^2, q_i q_z$ with $i=x,y$ as these contribute subleading terms in the $\omega$ dependence. We can perform the remaining integrals by scaling.  The step function sets $q_y\propto\omega^{1/2}$ and $q_z\propto\omega^{1/4}$ and hence
\begin{equation}
\tau^{-1}\propto \omega^{5/4}.
\end{equation}
This result can be derived from the golden-rule-type calculation in Eq.~\eqref{golden} as well.
The matrix element is dominated by $q_z^2\propto \omega^{2/4}$ and the phase space $q_yq_z$ scales as $\omega^{1/2}\omega^{1/4}$. Thus $1/\tau\propto \omega^{(2/4)+(1/2)+(1/4)}=\omega^{5/4}$.

Note that the exponent is larger than $1$, so that the quasiparticle excitation becomes sharper and sharper as we approach the Fermi surface ($\omega\rightarrow0$). Therefore, the single-particle excitation can be well described within Fermi-liquid theory.

The resistivity, on the other hand, exhibits an anomalous temperature dependence.  We introduce a $(1-\cos\theta_{\vec{k}',\vec{k}})$ factor in the above calculation to weight scattering processes in terms of their contribute to the transport relaxation time, which may then be evaluated as
\begin{eqnarray}
\tau_{\text{tr}}^{-1}&\simeq&\frac{V^i_zV^i_z}{4sB_0}\int\frac{\mathrm{d}^2q_{\parallel}\mathrm{d}q_z}{(2\pi)^3}q_z^2\theta(\omega-\epsilon_{\vec{q}})(1-\cos\theta_{\vec{k}',\vec{k}})\notag\\
&&\times\delta(\omega-(2k_Fq_x+q_y^2)/2m_{\parallel}-q_z^2/2m_z+\epsilon_{\vec{q}})\notag\\
&\simeq&\frac{V^i_zV^i_zm_{\parallel}}{4sB_0k_F^3}\int\frac{\mathrm{d}q_y\mathrm{d}q_z}{(2\pi)^2}q_z^4\theta\left(\omega-\frac{\rho_{\parallel}q_y^2+\rho_z q_z^4}{2sB_0}\right)\notag\\
&\propto&\omega^{7/4}.
\end{eqnarray}
As the exponent is lower than $2$ (the prediction of Fermi-liquid theory), transport measurements therefore reveal non-Fermi-liquid behavior of electrons.  An identical conclusion applies to the case $\vec{k}=k_F\hat{y}$, owing to the in-plane rotational invariance of the phonon and electron dispersion.

However, the properties of the system along the columnar axis ($z$ direction) are qualitatively different from the two perpendicular directions along which the crystal is ordered.  For $\vec{k}=k_F\hat{z}$, after performing the $q_z$ integral using the delta function, we are left with $q_x$ and $q_y$ integrals, both of which scale with $\omega^{1/2}$. It is easy to see that the quasiparticle lifetime and the transport relaxation time behave as $\tau^{-1}\propto\omega^{2}$ and $\tau_{\text{tr}}^{-1}\propto\omega^{3}$. (The latter should be replaced by $\omega^{2}$, which should be in general produced by other mechanism than those considered here.)  These corrections perfectly fit within standard Fermi-liquid theory.  

Knowing the conductivity along the principal axes should be sufficient to deduce that for along arbitrary axes, as we know all nonzero components of the conductivity tensor,
\begin{equation}
\sigma=\begin{pmatrix}
CT^{-7/4}&\sigma_{xy}&0\\
-\sigma_{xy}&CT^{-7/4}&0\\
0&0&C'T^{-2}
\end{pmatrix}.
\end{equation}
Here, we assumed that external magnetic field is along the $z$ axis and $\sigma_{xy}$ is the Hall conductance including the topological Hall effect.

\section{Discussion and Conclusions}
\label{Sec:4}
The mechanism above leads to a resistivity exponent ($\delta \rho\sim T^{1.75}$) which lies between the experimentally observed ($\delta \rho\sim T^{1.5}$) in the high-pressure nFL phase,  and the Fermi-liquid prediction ($\delta \rho\sim T^2$). Moreover, this mechanism predicts a phase, rather than a critical point, over which the resistivity deviates from the familiar Fermi-liquid form.  A key ingredient in this picture is the spontaneous generation of a 3D SkX in zero field in the high pressure phase. Related scenarios have been alluded to in earlier theoretical work \cite{BinzPRL,Binz2006, Rossler,FischerRosch, Tewari2006} that sought to explain the observed ``partial order" which overlaps with the nFL behavior. Also, recent experiments \cite{pfleiderer2013} have observed a large topological Hall effect in the high-pressure phase on application of weak magnetic fields, consistent with an incipient SkX at zero field. The soft phonons of orientation symmetry breaking may thus provide a link between the unusual magnetic properties observed at high pressure and the nFL transport. However, we caution that a comparison to the actual experiment needs to account for the effects of pinning to the underlying crystal axis, which sets an energy scale below which the soft phonon dispersion is converted into a more conventional quadratic one; Fermi-liquid behavior should be recovered in the quadratic regime. Furthermore, a more complete transport theory will incorporate the effects of disorder, Fermi surface geometry, etc. Finally, an additional caveat is that within our scenario, we would expect the electronic structure to be modified into minibands by the underlying SkX, an effect that should be apparent in gross transport properties on cooling the system.  However such an evolution is not apparent in the experimental data, pointing, perhaps, to a disordered or fluctuating SkX. 

To conclude, in this paper we have demonstrated that Berry phase fluctuations in general do not give rise to non-Fermi-liquid behavior, despite the formal similarity to a system of electrons coupled to a dynamical gauge field.  Scattering from SkX phonons may well have important quantitative effects, which may be interesting to observe in experiments; however they are not expected to qualitatively affect low-energy universal properties. We then proposed an alternative and new example of anomalous Fermi-liquid behavior, where the transport is non-Fermi-liquid-like but the quasiparticle description continues to hold. Here,  scattering of electrons by ultrasoft Goldstone excitations ($\epsilon_p\propto p^4$) in a columnar SkX which spontaneously breaks rotation symmetry leads to  a resistivity that scales with temperature as $T^{7/4}$. However, the vanishing vertex function for a small momentum transfer leaves the Goldstone mode undamped. Note that the non-Fermi-liquid behavior in this case arises due to the softness of the mode, rather than an unusual coupling: the electron-phonon vertex vanishes at zero momentum transfer, as we have proved it must. 

\section{Acknowledgements}
We thank T. Senthil, Oleg Tchernyshyov, Leo Radzihovsky, Steve Kivelson, Tomas Brauner, and Andrew Potter for useful discussions. We are particularly indebted to Max Metlitski for penetrating comments on an earlier version of this paper. A.V. is supported by NSF DMR 0645691. S.A.P. is supported by a Simons Postdoctoral Fellowship at UC Berkeley. S.R. acknowledges support from the DOE Office of Basic Energy Sciences, Contract No. DE-AC02-76SF00515, the John Templeton Foundation, and the Alfred P. Sloan Foundation.

\bibliography{SkXRefs}
\end{document}